\documentclass[preprintnumbers,nofootinbib,amsmath,amssymb,12pt]{revtex4}

\usepackage{amssymb}
\usepackage{amsmath}
\usepackage{epsfig}
\usepackage{graphicx}
\usepackage{here}
\usepackage{color}
\def\lapproxeq{\lower .7ex\hbox{$\;\stackrel{\textstyle <}{\sim}\;$}}
\def\gapproxeq{\lower .7ex\hbox{$\;\stackrel{\textstyle >}{\sim}\;$}}

\begin{document} 
%
%
%
\title{Michel parameters for $\tau$ decays $\tau \rightarrow l\nu\bar{\nu}~(l=e,~\mu)$
  in a general two Higgs doublet model with $\mu-\tau$ flavor violation}

\author{
Kazuhiro Tobe}
%
%
\affiliation{
  Department of Physics, Nagoya University, Nagoya, 464-8602, Japan}
\affiliation{Kobayashi-Maskawa Institute for the Origin of Particles and the Universe,
          Nagoya University,
      Nagoya, 464-8602, Japan}
%
\begin{abstract}
  In a general two Higgs doublet model (2HDM), the anomaly of muon anomalous magnetic
  moment (muon g-2) can be explained by $\mu-\tau$ flavor violating Yukawa couplings, motivated by the
  recent CMS excess in Higgs boson decay $h\rightarrow \mu \tau$.
  We study Michel parameters for $\tau$ decays $\tau \rightarrow l \nu \bar{\nu}~(l=e,~\mu)$
  in the 2HDM with the lepton flavor violation, and show that they can be sensitive to the
  flavor structure as well as the Lorentz and chiral structures of the model.
  We find that the correction to the Michel parameter $\xi_\mu$ in $\tau \rightarrow \mu \nu\bar{\nu}$
  is correlated to the contribution to the muon g-2,
  and it can be as large as $10^{-4}-10^{-2}$ in the parameter region
  where the $\mu-\tau$ flavor violating Yukawa
  couplings explain the muon g-2 anomaly. Therefore the precision measurement
  of the Michel parameter at the level of $10^{-4}-10^{-2}$ would significantly probe the interesting
  parameter space for the solution to the muon g-2 anomaly.
\end{abstract}
%
\maketitle
%
%
%

\section{Introduction}
The discovery of a Higgs boson at the Large Hadron Collider (LHC)~\cite{Aad:2012tfa,Chatrchyan:2012xdj} as well as the consistency of
almost all low energy experimental results show the remarkable success of the standard model (SM) of elementary particles.
On the other hand, the theoretical understanding of the Higgs sector is still poor.
There are no apparent theoretical reason that the Higgs sector has to have the simplest structure (one Higgs doublet) in contrast
to the matter sector which has three generation structure.
Therefore, the extended Higgs sector would deserve to be studied to make a deep understanding of the nature of Higgs sector.

One of simple extensions of the SM Higgs sector is a two Higgs doublet model
(2HDM)~\footnote{See a recent review~\cite{Branco:2011iw}.}, where one more Higgs doublet is
added into the SM. In a general 2HDM where both Higgs doublets couple to all fermions~\footnote{
  This is sometimes called type-III 2HDM. (See, for example, Refs.~\cite{Liu:1987ng,Aoki:2009ha,Hou:1991un,Atwood:1996vj}.)
  However, sometimes the type-III 2HDM is referred to as a different type
of 2HDM~\cite{Barger:1989fj}. To avoid confusion, we call it a general 2HDM.},
flavor violating phenomena mediated by the Higgs bosons are predicted~\cite{Bjorken:1977vt}.
Without any experimental supports, such a flavor violation beyond the SM has been considered to be
problematic~\cite{Glashow:1976nt,McWilliams:1980kj,Shanker:1981mj,Cheng:1987rs}.

However, the CMS collaboration has reported an excess in a flavor violating Higgs boson decay
$h\rightarrow \mu\tau$ at $\sqrt{s}=8$ TeV~\cite{Khachatryan:2015kon}, and the best fit value of the branching ratio is
\begin{eqnarray}
{\rm BR}(h\rightarrow \mu\tau)=(0.84^{+0.39}_{-0.37})~\%,
\end{eqnarray}
and the deviation from the SM prediction is $2.4\sigma$. Recently, the CMS collaboration reported a
result based on an integrated luminosity of 2.3 fb$^{-1}$ at $\sqrt{s}=13$ TeV and no excess is
observed~\cite{CMS:2016qvi}, but it is not sensitive enough to exclude the 8 TeV result.
The ATLAS collaboration has also reported their results~\cite{Aad:2015gha,Aad:2016blu} and the current best fit
value of the branching ratio is
\begin{eqnarray}
{\rm BR}(h\rightarrow \mu\tau)=(0.53\pm 0.51)~\%,
\end{eqnarray}
which is consistent with the SM prediction as well as the CMS result shown above.
Although the origin of the excess is not conclusive yet and more data are needed, the CMS excess in the flavor
violating Higgs boson decay becomes a good motivation to study the flavor violating phenomena predicted in the 2HDM and multi-Higgs doublet model~\footnote{
  For earlier works, see, for example, Refs.~\cite{Campos:2014zaa,Sierra:2014nqa,Heeck:2014qea,Crivellin:2015mga,deLima:2015pqa,Dorsner:2015mja}.
  The lepton flavor violation Higgs boson decays have been studied even before the CMS excess
  has been reported~\cite{Pilaftsis:1992st,Korner:1992zk,Diazcruz:1999xe,Assamagan:2002kf,Brignole:2003iv,
    Kanemura:2004cn,Arganda:2004bz,Kanemura:2005hr,Davidson:2010xv,Blankenburg:2012ex,Harnik:2012pb,Arhrib:2012ax,Arana-Catania:2013xma,Arganda:2014dta,Kopp:2014rva}.}.

In Refs.~\cite{Omura:2015nja,Omura:2015xcg}, we have shown a possibility that a general 2HDM with
$\mu-\tau$ flavor violation can explain both the CMS excess in $h \rightarrow \mu\tau$ and
an anomaly of muon anomalous magnetic moment (muon g-2) as reported, for example, by~\cite{Hagiwara:2011af}~\footnote{
  Similar results have been obtained by~\cite{Jegerlehner:2009ry,Davier:2010nc,Jegerlehner:2011ti}.},
\begin{eqnarray}
a_\mu^{\rm EXP}-a_\mu^{\rm SM}=(26.1\pm 8.0)\times 10^{-10}.
\end{eqnarray}
In the scenario where the 2HDM with $\mu-\tau$ flavor violation can resolve both anomalies,
we have studied some predictions and constraints in $\mu$ and $\tau$-physics~\cite{Omura:2015xcg}.
Especially we have found that the correction to the decay rate of $\tau \rightarrow \mu \nu\bar{\nu}$
is correlated to the correction to the muon g-2, and hence the precise measurement of the $\tau$ decay
$\tau \rightarrow l \nu\bar{\nu}~(l=e,~\mu)$ is important to probe the scenario.

In this paper, we study Michel parameters for $\tau$ decays $\tau\rightarrow l \nu \bar{\nu}~(l=e,~\mu)$
in a general 2HDM with the lepton flavor violation.
The Michel parameters in the leptonic decays $l\rightarrow l'\nu\bar{\nu}$ has been studied, for example, in
\cite{Michel,Bouchiat:1957zz,Kinoshita:1957zz,Kinoshita:1957zza,Scheck:1977yg,Fetscher:1986uj,Pich:1995vj,Pich:2013lsa,Kuno:1999jp},
and within the framework of the 2HDM~\cite{Haber:1978jt,McWilliams:1980kj,Stahl:1993yk,Logan:2009uf,Abe:2015oca,Chun:2016hzs}.
However, the effect of the lepton flavor violation on the Michel parameters has not been well studied. Therefore,
we analyze the corrections to the Michel parameters in $\tau$ decays $\tau \rightarrow l\nu\bar{\nu}~(l=e,~\mu)$ for the 2HDM
in the presence of the lepton flavor violation.
We stress that the precise measurement of the Michel parameters would have a sensitivity not only to the Lorentz and chiral structures
but also to the flavor structure of the new physics models.
Furthermore, we calculate the size of the corrections to the Michel parameters in the scenario
where the muon g-2 anomaly is explained by the $\mu-\tau$ flavor violation in the 2HDM, and show that it can be
as large as $10^{-4}-10^{-2}$.
We also find that there is an interesting correlation between 
the corrections to the Michel parameter $\xi_\mu$ in $\tau \rightarrow \mu \nu\bar{\nu}$ and the muon g-2,
independent of the value of ${\rm BR}(h\rightarrow \mu\tau)$.
Therefore, the precise measurement of the Michel parameter at the level of $10^{-4}-10^{-2}$ would significantly test the scenario.

This paper is organized as follows. In section II, we briefly review a general 2HDM. In section III, we study Michel parameters
for $\tau$ decays $\tau^-\rightarrow l^-\nu \bar{\nu}~(l=e,~\mu)$ in a general 2HDM with lepton flavor violation. Especially
we show that the Michel parameters can be sensitive to the flavor structure as well as the Lorentz and chiral structures of
the model. In section IV, we show the predicted values of the correction to the Michel parameter $\Delta \xi_\mu$
in the scenario where the muon g-2 anomaly can be explained by the $\mu-\tau$ flavor violation in the 2HDM.
In section V, we summarize our results.

\section{General two Higgs doublet model}
We briefly review a two Higgs doublet model. In a two Higgs doublet model,
both neutral components of Higgs doublets get vacuum expectation values (vevs) in general.
Taking a certain linear combination, we can always consider a basis
(so called Georgi basis or Higgs basis~\cite{Georgi:1978ri,Donoghue:1978cj},
and see also, for example, \cite{Lavoura:1994fv,
  Lavoura:1994yu,Botella:1994cs,Branco:1999fs,Davidson:2005cw})
where only one of the Higgs doublets has the vev as follows:
\begin{eqnarray}
  H_1 =\left(
  \begin{array}{c}
    G^+\\
    \frac{v+\phi_1+iG}{\sqrt{2}}
  \end{array}
  \right),~~~
  H_2=\left(
  \begin{array}{c}
    H^+\\
    \frac{\phi_2+iA}{\sqrt{2}}
  \end{array}
  \right),
\label{HiggsBasis}
\end{eqnarray}
where $G^+$ and $G$ are Nambu-Goldstone bosons, and $H^+$ and $A$ are a charged Higgs boson and a CP-odd
Higgs boson, respectively. We have assumed that the CP is conserved in the Higgs potential for simplicity.
CP-even neutral Higgs bosons $\phi_1$ and $\phi_2$ can mix and form mass eigenstates, $h$ and $H$ ($m_H>m_h$),
\begin{eqnarray}
  \left(
  \begin{array}{c}
    \phi_1\\
    \phi_2
  \end{array}
  \right)=\left(
  \begin{array}{cc}
    \cos\theta_{\beta \alpha} & \sin\theta_{\beta \alpha}\\
    -\sin\theta_{\beta \alpha} & \cos\theta_{\beta \alpha}
  \end{array}
  \right)\left(
  \begin{array}{c}
    H\\
    h
  \end{array}
  \right).
\end{eqnarray}
Here $\theta_{\beta \alpha}$ is the mixing angle.

Without imposing an extra symmetry, both Higgs doublets couple to all fermions.
In mass eigenbasis for the fermions, the lepton Yukawa interactions are expressed by
\begin{eqnarray}
  {\cal L} &=& -\bar{L}_L^i H_1 y^i_e e_R^i -\bar{L}_L^i H_2 \rho^{ij}_e e_R^j+{\rm h.c.},
\end{eqnarray}
where $i,j$ represent flavor indices, $L_L=(V_{\rm MNS} \nu_L, e_L)^T$, and 
$V_{\rm MNS}$ is the Maki-Nakagawa-Sakata (MNS) matrix. Here all fermions
$(f_L,~f_R)$ $(f=e,\nu)$ are mass eigenstates
(i.e. $e_{L,R}^1=e_{L,R},~e_{L,R}^2=\mu_{L,R},~e_{L,R}^3=\tau_{L,R}$). We have assumed the seesaw mechanism
with super-heavy right-handed neutrinos to explain the smallness of neutrino masses.
 The Yukawa coupling matrix $\rho_e^{ij}$ is a general $3\times3$
complex matrix and can be a source of the Higgs-mediated flavor violating processes.
Although we only show Yukawa couplings in lepton sector, the Yukawa couplings in quark sector
are understood similarly.

In mass eigenstates of Higgs bosons, the lepton Yukawa interactions are given by
\begin{align}
  {\cal L}=-\sum_{\phi=h,H,A} y^e_{\phi i j}\bar{e}_{Li} \phi e_{Rj} 
-\bar{\nu}_{Li} (V_{\rm MNS}^\dagger \rho_e)^{ij}  H^+ e_{Rj}+{\rm h.c.},
\end{align}
where
\begin{align}
  y^e_{hij}=\frac{m_{e}^i}{v}s_{\beta\alpha}\delta_{ij}+\frac{\rho_{e}^{ij}}{\sqrt{2}} c_{\beta\alpha},~~~
  y^e_{Hij}=\frac{m_e^i}{v} c_{\beta \alpha}\delta_{ij}-\frac{\rho_e^{ij}}{\sqrt{2}} s_{\beta\alpha},~~~
  y^e_{Aij}=
    \frac{i\rho_e^{ij}}{\sqrt{2}},
  \label{yukawa}
\end{align}
where $c_{\beta\alpha}\equiv \cos\theta_{\beta\alpha}$ and $s_{\beta\alpha}\equiv\sin\theta_{\beta\alpha}$,
and $m_e^i=y_e^i v /\sqrt{2}$.
Note that when $c_{\beta\alpha}=0~(s_{\beta\alpha}=1)$, the Yukawa interactions of $h$ are equal to those of
the SM Higgs boson. In general, however, there are flavor-violating interactions for $h$ through the Higgs
mixing $c_{\beta\alpha}$. On the other hand, when $c_{\beta\alpha}$ is small,
the Yukawa interactions of heavy Higgs bosons ($H$, $A$, and $H^+$) mainly come from the $\rho_e$
couplings.

A scalar potential in the general 2HDM is given by
\begin{align}
  V&=M_{11}^2 H_1^\dagger H_1+M_{22}^2 H_2^\dagger H_2 \nonumber -\left(M_{12}^2H_1^\dagger H_2+{\rm h.c.}
  \right)\\
  &+\frac{\lambda_1}{2}(H_1^\dagger H_1)^2 \nonumber
  +\frac{\lambda_2}{2}(H_2^\dagger H_2)^2+\lambda_3(H_1^\dagger H_1)(H_2^\dagger H_2)
  +\lambda_4 (H_1^\dagger H_2)(H_2^\dagger H_1)\\
  &+ \frac{\lambda_5}{2}(H_1^\dagger H_2)^2\nonumber
  +\left\{
\lambda_6 (H_1^\dagger H_1)+\lambda_7 (H_2^\dagger H_2)\right\} (H_1^\dagger H_2)+{\rm h.c.}.
\end{align}
From this potential, one can calculate the relations among Higgs boson masses,
and especially when $c_{\beta \alpha}$ is close to zero (or $\lambda_6 \sim 0$), the 
relations are simplified as
\begin{align}
  m_h^2& \simeq \lambda_1 v^2,~~~  m_H^2 \simeq m_A^2+\lambda_5 v^2,\nonumber \\
  m_{H^+}^2 &= m_A^2-\frac{\lambda_4-\lambda_5}{2} v^2,~~~ m_A^2  = M_{22}^2+\frac{\lambda_3+\lambda_4-\lambda_5}{2} v^2.
  \label{Higgs_spectrum2}
\end{align}
Fixing the couplings $\lambda_i$, the heavy Higgs boson masses
are parametrized by the CP-odd Higgs boson mass $m_A$, which we treat as a free
parameter of the model.
We note that a dangerous contribution to Peskin-Takeuchi's T-parameter ($\rho$ parameter)~\cite{peskin}
are suppressed by the degeneracy between $m_A$ and $m_{H^+}$ as well as the small Higgs mixing
parameter $c_{\beta \alpha}$~\cite{Haber:2010bw}. Therefore, we set $\lambda_4=\lambda_5=0.5$ in our analysis, so that
it guarantees the degeneracy between the CP-odd Higgs and charged Higgs bosons $m_A=m_{H^+}$~\footnote{
  In this Higgs boson mass spectrum, Peskin-Takeuchi's S and U parameters are also small~\cite{Haber:2010bw}.}.

\section{Michel parameters for $\tau$ decays $\tau^- \rightarrow l^- \nu \bar{\nu}~(l=e,~\mu)$}
In the 2HDM, charged Higgs boson interactions also induce $\tau$ decays $\tau^-\rightarrow l^- \nu \bar{\nu}$
at the tree level. Therefore, the detail study of the $\tau$ decays is interesting to see the new physics effect.
For an initial $\tau^-$ lepton polarization $\cal{P}_\tau$, the final $l^-$ distribution $(l=e,~\mu)$
in the $\tau$ rest frame of $\tau^-\rightarrow l^- \nu \bar{\nu}$ decay
is given in terms of Michel parameters $\rho_l,~\eta_l,~\xi_l$ and $\delta_l$\cite{
  Michel,Bouchiat:1957zz,Kinoshita:1957zz,Kinoshita:1957zza,Scheck:1977yg,Fetscher:1986uj,Pich:1995vj,Pich:2013lsa,Kuno:1999jp}:
\begin{eqnarray}
\frac{d\Gamma (\tau^-\rightarrow l^- \nu \bar{\nu})}{dx d\cos\theta_l}&=&
  \frac{m_\tau w^4}{2\pi^3}\sqrt{x^2-x_0^2}G_{F_l}^2
  \left[F_{1 l}(x)-F_{2 l}(x){\cal{P}}_\tau \cos\theta_l \right],\nonumber \\
    F_{1 l}(x)&=&x(1-x)+\frac{2\rho_l}{9}(4x^2-3x-x_0^2)
  +\eta_l x_0 (1-x),\nonumber \\
  F_{2 l}(x)&=&
  \frac{\xi_l \sqrt{x^2-x_0^2}}{3}
  \left\{ 1-x+
  \frac{2\delta_l(4x-4+\sqrt{1-x_0^2})}{3}\right\},
\end{eqnarray}
where $G_{F_l}$ is an effective Fermi constant for $\tau^- \rightarrow l^- \nu \bar{\nu}$
process, and 
$\theta_l$ is the angle between the $\tau^-$ spin and the final $l^-$ momentum, 
$w$ is the maximum $l^-$ energy ($w=\frac{m_\tau^2+m_l^2}{2m_\tau}$), and
$x=E_l/w$ and $x_0=m_l/w$ where $E_l$ and $m_l$ are
energy and mass for the lepton $l$ ($l=e,\mu$), respectively.
Here we have assumed
neutrino masses are negligible. 
The decay rate for $\tau^-\rightarrow \l^- \nu\bar{\nu}$ is expressed by
\begin{eqnarray}
  \Gamma_l&=&
  \frac{G_{F_l}^2m_\tau^5}{192 \pi^3}\left\{
  f(y_l)+4\eta \frac{m_l}{m_\tau} g(y_l) \right\},
  \label{Gamma}
\end{eqnarray}
where $y_l=\frac{m_l^2}{m_\tau^2}$, $f(y)=1-8y+8y^3-y^4-12y^2\log y$, and
$g(y)=1+9y-9y^2-y^3+6y(1+y)\log y$.

In the 2HDM, the effective Fermi constant $G_{F_l}$ and the  Michel parameters
for $\tau^-\rightarrow l^- \nu \bar{\nu}$ are expressed by
\begin{eqnarray}
  G_{F_l} = G_F \sqrt{1+\Delta^l_1},~~~
  \rho_l = \frac{3}{4},~\delta_l=\frac{3}{4},~~~
  \xi_l = \frac{1-\Delta^l_1}{1+\Delta^l_1},~~~
  \eta_l = -\frac{\Delta^l_2}{1+\Delta^l_1}.
\end{eqnarray}
Here the corrections $\Delta^l_1$ and $\Delta^l_2$
are defined by
\begin{eqnarray}
  \Delta^l_1 =
  \frac{(\rho_e^\dagger \rho_e)^{ll} (\rho_e^\dagger \rho_e)^{\tau\tau}}{32 G_F^2 m_{H^+}^4},~~~
  \Delta^l_2 = \frac{{\rm Re}(\rho_e^{ll}\rho_e^{\tau\tau*})}{4\sqrt{2} G_F m_{H^+}^2},
\end{eqnarray}
where $m_{H^+}$ is the charged Higgs boson mass. Since the flavor of neutrinos and anti-neutrinos are not
detected in the measurement, we have taken a sum of the flavor of neutrinos and anti-neutrinos in the final state.
Thus we expect the deviation from the SM prediction in $\xi_l$ and $\eta_l$,
\begin{eqnarray}
  \Delta \xi_l= \xi_l-\xi_{\rm SM}=-\frac{2\Delta_1^l}{1+\Delta_1^l}\simeq -2\Delta_1^l,\\
  \Delta \eta_l =\eta_l -\eta_{\rm SM}=-\frac{\Delta_2^l}{1+\Delta_1^l}\simeq -\Delta_2^l,
\end{eqnarray}
where $\xi_{\rm SM}=1$ and $\eta_{\rm SM}=0$ for the standard model values.

We note that if there are only flavor-conserving interactions assuming CP conservation for simplicity,
the $\Delta_1^l$ and $\Delta_2^l$ are related:
\begin{eqnarray}
\Delta_1^l=\frac{(\rho_e^{ll} \rho_e^{\tau\tau})^2}{32 G_F^2 m_H^4}=(\Delta_2^l)^2,
\end{eqnarray}
and hence we expect that $|\Delta \eta_l|\gg |\Delta \xi_l|$. On the other hand, if the flavor-violating interactions
are dominant, the relation between $\Delta \eta_l$ and $\Delta \xi_l$ would be very different from the one
in the flavor-conserving case. For example, if only $\rho_e^{\mu \tau~(\tau \mu)}$ are non-zero and others are
negligible,
\begin{eqnarray}
  \Delta_1^e \simeq 0,~~~\Delta_1^\mu \simeq \frac{|\rho_e^{\tau \mu} \rho_e^{\mu\tau}|^2}{32 G_F^2 m_H^4},~~~
  \Delta_2^l \simeq  0~{\rm for}~l=e~{\rm and}~\mu,
  \label{example1}
\end{eqnarray}
so that $|\Delta \xi_\mu| \gg |\Delta \eta_\mu|$~\footnote{When the $\rho^{\mu\tau~(\tau\mu)}_e$ flavor
  violating Yukawa couplings are dominant, the flavors of neutrino and anti-neutrino in the final
  state are different from those of the SM contribution. Therefore, there is no interference between the SM and charged Higgs contributions.}.
Therefore we stress that the precise measurement of various
Michel parameters are very important to understand not only the Lorentz and chiral structure but also
the flavor structure of the new physics models.

Experiments have performed a test of lepton flavor universality by measuring the following
quantity:
\begin{eqnarray}
  \left(\frac{g_\mu}{g_e}\right)_\tau^2
  \equiv
  \frac{{\rm BR}(\tau^-\rightarrow \mu^-\bar{\nu}\nu) f(y_e)}
       {{\rm BR}(\tau^- \rightarrow e^- \bar{\nu}\nu) f(y_\mu)},
\end{eqnarray}
where $f(y)$ is the same function shown in Eq.(\ref{Gamma}).
The current world average~\cite{Aubert:2009qj} is
\begin{eqnarray}
  \left(\frac{g_\mu}{g_e}\right)_\tau
  =1.0018\pm 0.0014.
  \label{exp_limit}
\end{eqnarray}
In the 2HDM, this quantity is given by
\begin{eqnarray}
  \left(\frac{g_\mu}{g_e}\right)_\tau^2 =
  \frac{G^2_{F_\mu}}{G^2_{F_e}}
  \frac{1+4\eta_\mu \frac{m_\mu}{m_\tau}\gamma(y_\mu)}
       {1+4\eta_e \frac{m_e}{m_\tau} \gamma(y_e)}
       =\left(\frac{1+\Delta_1^\mu}{1+\Delta_1^e}\right)\left(
         \frac{1+4\eta_\mu \frac{m_\mu}{m_\tau}\gamma(y_\mu)}
       {1+4\eta_e \frac{m_e}{m_\tau} \gamma(y_e)}\right),
\end{eqnarray}
where $\gamma(y_l)=g(y_l)/f(y_l)$.
Therefore, the measurement of the lepton flavor universality is sensitive to the
non-universality of the effective Fermi constant $G_{F_l}$ (in other word, $\Delta_1^l$)
as well as the parameter $\eta_l$ $(\Delta_2^l)$. Especially, in the case with negligible $\eta_l~(\Delta_2^l)$
as shown in Eq.~(\ref{example1}), the correction to the lepton non-universality
is sensitive to the lepton flavor violation~\cite{Omura:2015xcg} and it is related to
the correction to the Michel parameter $\xi_\mu$;
\begin{eqnarray}
  \left(\frac{g_\mu}{g_e}\right)_\tau \simeq 1+\frac{\Delta_1^\mu}{2}\simeq 1-\frac{\Delta \xi_\mu}{4}.
  \label{non-univ}
\end{eqnarray}
Since $\Delta \xi_\mu<0$, $(g_\mu/g_e)_\tau>1$ in this scenario.

\section{Correlation between corrections to muon g-2 and Michel parameter $\xi_\mu$}
In Refs.~\cite{Omura:2015nja, Omura:2015xcg}, we have found that the anomaly of muon g-2 can
be explained by $\mu-\tau$ flavor violating Yukawa interactions in a general 2HDM,
which is motivated by the CMS excess in the Higgs boson decay $h\rightarrow \mu \tau$~\cite{Khachatryan:2015kon}.
It will be interesting to see how large correction to the Michel parameters one can get in the parameter
space where the muon g-2 anomaly is explained.

\begin{figure}[t]
  \begin{center}%
    \epsfig{figure=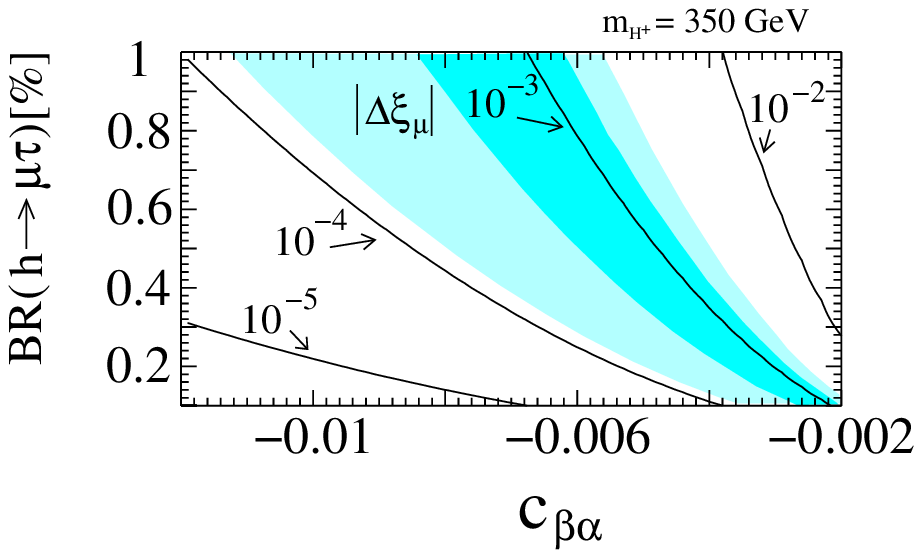,width=0.8\textwidth}
    \epsfig{figure=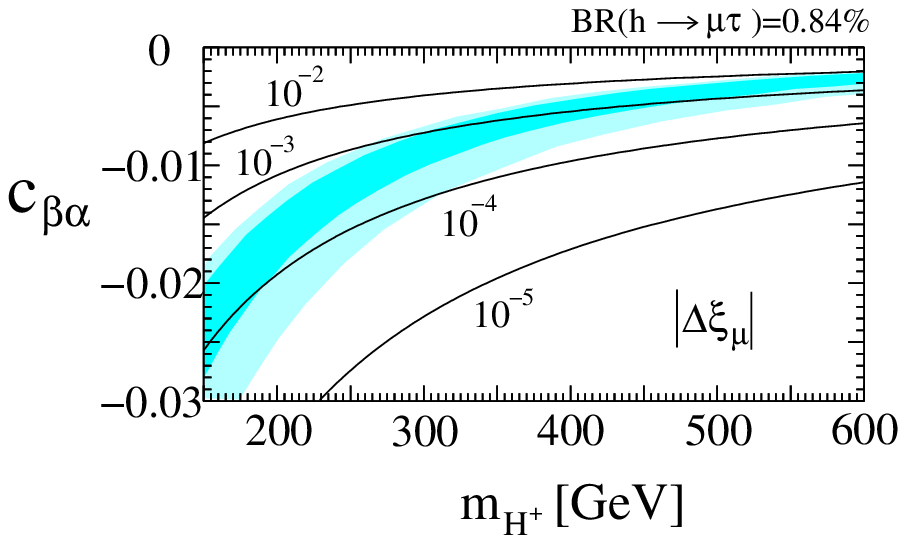,width=0.82\textwidth}
    \caption{The correction to the Michel parameter $|\Delta \xi_\mu|$ is shown as a function of $c_{\beta \alpha}$ and
      ${\rm BR}(h \rightarrow \mu \tau)$ (upper figure) and as a function of
      charged Higgs boson mass $m_{H^+}$ and $c_{\beta \alpha}$ (lower figure). 
      We have assumed $m_{H^+}=350$ GeV in the upper figure and ${\rm BR}(h\rightarrow \mu\tau)=0.84\%$ in the lower figure.
      The dark (light) shaded region can resolve the muon g-2 anomaly within $\pm 1\sigma~(\pm 2\sigma)$.}
    \label{michel_para}
  \end{center}
\end{figure}

In an upper figure of Fig.~{\ref{michel_para}}, we show the
absolute value of the correction to the Michel parameter $|\Delta \xi_\mu|$ as a function of $c_{\beta \alpha}$ and
${\rm BR}(h \rightarrow \mu \tau)$. Here we have assumed $m_{H^+}=350$ GeV.
In a lower figure of Fig.~{\ref{michel_para}},
$|\Delta \xi_\mu|$ is shown as a function of charged Higgs boson mass $m_{H^+}$ and $c_{\beta \alpha}$ fixing the branching
ratio ${\rm BR}(h\rightarrow \mu \tau)~({\rm BR}(h\rightarrow \mu \tau)=0.84\%)$.
The dark (light) shaded region can explain the muon g-2 anomaly within $\pm 1 \sigma~(\pm 2\sigma)$.
In these figures, we have assumed that only flavor violating Yukawa couplings $\rho^{\mu\tau~(\tau\mu)}_e$
are non-zero, and others $\rho_e$ Yukawa couplings are
negligible as we have discussed in Eq.~(\ref{example1})~\footnote{
  As shown in Ref.~{\cite{Omura:2015xcg}}, many of $\rho_e$ Yukawa couplings are
  strongly constrained in this scenario. Therefore, we simply neglect the others to focus on the effect
  of $\rho_e^{\mu\tau~(\tau\mu)}$ couplings.}. In order to explain the muon g-2 anomaly and to maximize its size,
we have assumed
  $\rho_e^{\mu\tau}=-\rho_e^{\tau\mu}$, as discussed in Ref.~\cite{Omura:2015xcg}.
As shown in Eq.~(\ref{example1}), $\Delta_1^\mu$ is always positive and hence $\Delta \xi_\mu$ is
negative.
As one can see from the upper figure of Fig.~{\ref{michel_para}}, there is an interesting correlation between
the corrections to the muon g-2 and the Michel parameter $\Delta \xi_\mu$ in $\tau\rightarrow \mu \nu\bar{\nu}$ decay,
that is almost independent of the value of ${\rm BR}(h\rightarrow \mu\tau)$.
This is in contrast to the prediction of $\tau \rightarrow \mu \gamma$ which depends on the value of
${\rm BR}(h\rightarrow \mu\tau)$~\cite{Omura:2015xcg}.
In the lower figure of Fig.~\ref{michel_para}, as the charged Higgs boson gets heavier, the predicted correction to
the Michel parameter $|\Delta\xi_\mu|$ becomes larger in the parameter region where the muon g-2 anomaly is
explained. The accuracy of the current Michel parameter measurements is at the $O(1)\%$ level~\cite{Agashe:2014kda}, and hence the results
are consistent with the current measurements on the Michel parameters.
Since the correction $|\Delta \xi_\mu|$ and the lepton non-universality $(g_\mu/g_e)_\tau-1$ in the
$\tau$ decays are related as shown in Eq.~(\ref{non-univ}), the current bound of the lepton non-universality
Eq.~(\ref{exp_limit}) starts putting on the constraint in this scenario. Therefore, the future precise measurement
of the Michel parameter $\xi_\mu$ at the level of $10^{-4}-10^{-2}$ as well as that of the lepton non-universality
would have a significant potential to probe this scenario.

\section{Summary}
The theoretical understanding of the Higgs sector is still unsatisfactory. The more experimental data and theoretical
studies will be needed to make a deeper understanding of the Higgs sector.

The CMS excess events of $h \rightarrow \mu\tau$ process
might suggest the extension of the minimal structure of the SM Higgs sector.
One of simple extensions of the SM Higgs sector is a two Higgs doublet extension of the SM. In a general 2HDM, the flavor
violating phenomena mediated by Higgs bosons are predicted, and hence it is easy to explain the CMS excess in $h\rightarrow
\mu\tau$ if this is due to new physics effect. In Refs.~\cite{Omura:2015nja,Omura:2015xcg}, we have pointed out that
the $\mu-\tau$ flavor violating Yukawa interactions can resolve the muon g-2 anomaly, and in Ref.~\cite{Omura:2015xcg},
the correction to the decay rate of $\tau\rightarrow \mu\nu\bar{\nu}$ process is correlated to the contribution to the muon g-2 induced by
the $\mu-\tau$ lepton flavor violating Yukawa interactions.

In this paper, we have studied the Michel parameters for the $\tau$ decays $\tau^-\rightarrow l^- \nu\bar{\nu}$
in a general 2HDM with lepton flavor violation, whose effect on the Michel parameters had not been well studied.
We have shown that the precise measurement of the Michel parameters is sensitive to the flavor structure as well as
the Lorentz and chiral structure of the model. Especially in the parameter region where the muon g-2 anomaly is
explained by the $\mu-\tau$ flavor violating Yukawa couplings, the correction to the Michel parameter $|\Delta \xi_\mu|$
can be as large as $10^{-4}-10^{-2}$ and it is correlated to the correction to the muon g-2, independent of the predicted
value of ${\rm BR}(h\rightarrow \mu\tau)$.
Therefore, the precision measurement of the Michel parameters at the level of $10^{-4}-10^{-2}$ would be crucial to
probe the scenario where the $\mu-\tau$ flavor violating Yukawa couplings explain the anomaly of the muon g-2.

\section*{Acknowledgments}
The author would like to thank Kiyoshi Hayasaka for suggesting him a study of Michel parameters in $\tau$ decays.
He would also like to thank James D. Wells and C.-P. Yuan for useful discussions during his visits
to University of Michigan and Michigan State University, respectively.
This work was supported in part by Japan Society for Promotion of Science (JSPS)  (No.26104705 and No.16K05319).

\end{document}